\documentclass[twocolumn,trackchanges]{aastex701}

\accepted{to ApJ}

\shorttitle{}
\shortauthors{Bennet \& Patel et al.}

\defcitealias{Bennet_2024}{B24}
\begin{document}

\title{The Orbits of Isolated Dwarfs in the Local Group from New 3D Kinematics: Constraints on First Infall, Backsplash, and Quenching Mechanisms}

\author[0000-0001-8354-7279]{Paul Bennet}\thanks{Patel and Bennet are co-first authors and have contributed \newline equally to the analysis and preparation of this manuscript.}
\affiliation{Space Telescope Science Institute, 3700 San Martin Drive, Baltimore, MD 21218, USA}
\email{pbennet@stsci.edu}

\author[0000-0002-9820-1219]{Ekta~Patel}\thanks{NASA Hubble Fellow}
\affiliation{Department of Physics and Astronomy, University of Utah, 115 South 1400 East, Salt Lake City, Utah 84112, USA}
\affiliation{Department of Astrophysics and Planetary Sciences, Villanova University,  800 E. Lancaster Ave, Villanova, PA 19085, USA}
\email{ekta.patel@villanova.edu}

\correspondingauthor{Paul Bennet}
\email{pbennet@stsci.edu}

\author[0000-0001-8368-0221]{Sangmo Tony Sohn}
\affiliation{Space Telescope Science Institute, 3700 San Martin Drive, Baltimore, MD 21218, USA}
\affiliation{Department of Astronomy \& Space Science, Kyung Hee University, 1732 Deogyeong-daero, Yongin-si, Gyeonggi-do 17104, Republic of Korea}
\email{tsohn@stsci.edu}

\author[0000-0003-4922-5131]{Andr\'es del Pino}
\affiliation{Instituto de Astrof\'isica de Andaluc\'ia, IAA-CSIC, Glorieta de la Astronom\'ia s/n, 18008, Granada, Spain}
\email{apino@iaa.csic.es}

\author[0000-0001-7827-7825]{Roeland P. van der Marel}
\affiliation{Space Telescope Science Institute, 3700 San Martin Drive, Baltimore, MD 21218, USA}
\affiliation{Center for Astrophysical Sciences, The William H. Miller III Department of Physics \& Astronomy, Johns Hopkins University, Baltimore, MD 21218, USA}
\email{marel@stsci.edu}

\author[0000-0003-4207-3788]{Mark A.\ Fardal}
\affiliation{Eureka Scientific, 2452 Delmer Street, Suite 100, Oakland, CA 94602, USA}
\email{mfardal@eurekasci.com}

\author[0000-0002-0956-7949]{Kristine Spekkens}
\affiliation{Department of Physics, Engineering Physics and Astronomy, Queen’s University, Kingston, ON K7L 3N6, Canada}
\email{kristine.spekkens@gmail.com}

\author[0000-0001-5368-3632]{Laura Congreve Hunter}
\affiliation{Department of Physics and Astronomy, Dartmouth College, 6127 Wilder Laboratory, Hanover, NH 03755, USA}
\email{Laura.C.Hunter@dartmouth.edu}

\author[0000-0003-0715-2173]{Gurtina~Besla}
\affiliation{Steward Observatory, University of Arizona, 933 North Cherry Avenue, Tucson, AZ 85721, USA}
\email{gbesla@arizona.edu}

\author[0000-0002-1343-134X]{Laura L. Watkins}
\affiliation{AURA for the European Space Agency (ESA), ESA Office, Space Telescope Science Institute, 3700 San Martin Drive, Baltimore, MD 21218, USA}
\email{lwatkins@stsci.edu}

\author[0000-0002-6442-6030]{Daniel~R.~Weisz}\affil{Department of Astronomy, University of California, Berkeley, CA 94720-3411, USA}
\email{drweisz@berkeley.edu}

\begin{abstract}
It is commonly supposed that quenched field dwarfs near the edge of the Local Group (LG) are backsplash galaxies, having previously orbited within the MW or M31’s virial radius, whereas galaxies on first infall should still have gas and star formation. We measured proper motions (PMs) for six dwarf galaxies located 400–1000 kpc from the Milky Way (MW) using the \textit{Hubble Space Telescope} (HST). For four galaxies (Aquarius, Cetus, Pisces, Tucana), we report the first PMs. For the remaining two (Leo T and Pegasus), we measure PMs with order-of-magnitude precision improvement. We compute orbital histories to assess whether any of the six are backsplash galaxies. While some have non-zero likelihoods of past interaction with the MW or M31, these are weak and typically occur at large distances (i.e., $> 2 R_{\rm vir}$). The properties of Aquarius, Leo T, Pisces, and Pegasus are consistent with first passage through a massive halo. Cetus, which shows a low probability ($\sim$4–6\%) of interacting with the MW or M31 in the last 6 Gyr, is more likely a backsplash galaxy resulting from an interaction with M31 over 6 Gyr ago, in the regime where rigid orbital models become less reliable. Tucana has been thought to be a backsplash galaxy, but our orbits indicate it cannot have interacted with a massive LG host. Our results highlight the diversity of evolutionary pathways for isolated, intermediate-mass dwarfs ($M_* \approx 10^5–10^7 \, \rm M_{\odot}$) and the need to reassess quenching mechanisms beyond environmental interactions with massive hosts.
\end{abstract}


\section{Introduction} \label{sec:intro}

The link between star formation and quenching is a key area of study in the dwarf galaxy regime, both in observations \citep[][]{Geha_2012,Karunakaran_2021,Karunakaran_2023,Font_2022,Sales_2022, Jones2023} and simulations \citep[e.g.][]{Fattahi_2016,Wetzel_2016,Simpson_2018,Garrison-Kimmel2019,Libeskind_2020,Samuel_2022,Gutcke_2022,Engler_2023}. In the Local Group (LG), most dwarf galaxies within the virial radius of the massive galaxies---Andromeda (M31) or the Milky Way (MW)---are quenched except for the most massive dwarfs, such as the Magellanic Clouds. On the other hand, isolated or field dwarfs---those outside the virial radius of a massive galaxy---are generally gas-rich and star forming \citep[][]{Spekkens_2014,Putman_2021}, especially for dwarfs with stellar masses $M_* > 10^9 \,\rm  M_{\odot}$ at more than 1.5 Mpc from a more massive galaxy \citep{Geha_2012}. 
This trend is also seen in simulations of LG-like galaxy groups and isolated MW-like galaxies \citep[e.g.][]{Fattahi_2016,Sawala_2016,Wetzel_2016,Simpson_2018,Garrison-Kimmel2019,Libeskind_2020,Engler_2021,Engler_2023,Font_2021,Applebaum_2021,Akins_2021,Samuel_2022}{}{}. 

Isolated dwarfs are key to solving questions surrounding star formation and quenching, as it is difficult to disentangle the effects of internal (e.g., reionization and supernova feedback) and external (tidal and ram pressure stripping) processes on dwarf galaxies in the vicinity of massive galaxies \citep[][]{Gatto_2013,Emerick_2016, Tollerud_2018, Putman_2021}. The influence of massive hosts can also extend to dwarfs far past their virial radius. Such galaxies are often referred to as backsplash galaxies. These are galaxies that are currently outside the virial radius of a massive host, but previously passed inside that radius and therefore may carry the imprints of any past interactions \citep[e.g][]{Gill_2005,Sales_2007,Fraternali_2009,Santos_santos_2023}. Such imprints include a lack of a significant HI gas reservoir, a lack of recent star formation, or morphological asymmetries.

Distinguishing galaxies that are backsplash from those that are on first infall requires 3D position and velocity information (together referred to as 6D phase space information) to reconstruct plausible orbital histories for these objects. However, such 6D phase space information is challenging to obtain. Historically, almost all of our knowledge of LG galaxy orbits has come from line-of-sight (LOS) velocities ($ V_{los}$). These motions are limited to motions towards or away from us, thus severely limiting characterization of wider dynamical properties and permitting a wide range of possible orbits. 
Proper motions (PMs), motions on the plane of the sky, provide the additional two components of motion needed for complete 6D characterization of a galaxy's kinematics and position, and therefore more complex kinematic properties, e.g., orbits. With sufficiently accurate PMs, we can use these derived orbits to distinguish between first infall and backsplash galaxies \citep[e.g.,][]{Bennet_2024}. 

In recent years PMs from \textit{Gaia} have tightly constrained the 3D motions of M31 and M33 \citep{vanderMarel2019, Salomon2021, Rusterucci_2024}, and have transformed our understanding of the MW satellites \citep{Simon2018, Fritz2018, Vasiliev2019, Patel_2020, Battaglia_2022}. 
However, motions of dwarfs outside of the outer MW halo are still mainly limited to one dimension. \textit{Gaia}'s present precision (DR3) is not sufficient for measuring bulk PMs of low-mass systems beyond $\sim$400~kpc \citep{Fritz2018}, and its magnitude limit at $G \sim 20.7$~mag means that some more distant objects will remain permanently out of reach, even with future data releases\footnote{The current DR3 has a time baseline of 34 months, however even the final 10 year catalog (DR5) will not improve on this photometric limit.}. 

Efforts have been made to combine the \textit{Hubble Space Telescope} (\textit{HST}) and \textit{Gaia} to get around some of these shortcomings \citep[][]{Massari_2020,del_Pino_2022,Bennet_2022,McKinnon_2023, Warfield_2023}. However, these efforts are also limited to the areas where \textit{Gaia} provides access to stellar populations that are bright enough and statistically significant, limiting it to nearby and star-forming objects, which often have upper main sequence member stars bright enough to appear in the \textit{Gaia} catalog even beyond the MW halo. Efforts are also underway to combine \textit{Gaia} with \textit{Euclid} \citep[][]{Libralato_2024}, though this has the same issues as combining \textit{HST} and \textit{Gaia}. 

Meanwhile, \textit{HST} can measure the PMs of even the most distant of LG galaxies, or even potentially beyond the LG, when two or more epochs of observations are available with sufficient temporal separation (time baseline) between them \citep[e.g.][]{Sohn_2013,Sohn_2017,Sohn_2020}. Here we report the PMs of six distant isolated LG dwarf galaxies using \textit{HST}. Our sample members were chosen due to their isolated status and lack of constraining PMs in the literature, with only two of the six having any measurements, albeit with highly-uncertain tangential velocities \citep[][]{McConnachie_2020a,McConnachie_2020b,Battaglia2022}. These two galaxies have also been examined through a combination of \textit{Gaia} and \textit{HST} \citep[][hereafter B24]{Bennet_2024}, and while those uncertainties improved on the \textit{Gaia}-only results, they are still not constrained sufficiently to provide physically-meaningful orbits. 

This paper is organized as follows. 
In \S \ref {sec:data}, we derive the new PMs for our dwarf sample and compare them to existing literature PMs. In \S \ref{sec:Orbits}, we use the newly-derived PMs, combined with literature positions, distances and LOS velocities, to derive possible orbital histories for our sample of LG dwarfs. In \S \ref{sec:discussion}, we discuss the implications of these results for each galaxy and compare their orbital and star formation histories. We also determine which dwarfs in the sample are potential backsplash galaxies and which are on their first passage through the halo of the MW or M31. Finally, in \S \ref{sec:conclusion}, we summarize our work and present our findings.

\section{Data} \label{sec:data}
\begin{deluxetable*}{lcccccccc}[t]
\tablecaption{Summary of imaging epochs
              \label{tab:epochs}}
\tablehead{
\colhead{}       & \multicolumn{2}{c}{{\bf Epoch~1}}    & \colhead{} & \multicolumn{2}{c}{{\bf Epoch~2}}    & \colhead{} & \multicolumn{2}{c}{{\bf Epoch~3}}   \\
\cline{2-3}\cline{5-6}\cline{8-9}
\colhead{Galaxy}       & \colhead{Date}    & \colhead{Exp. Time}      & \colhead{} & \colhead{Date}    & \colhead{Exp. Time}     & \colhead{} & \colhead{Date}    & \colhead{Exp. Time}}
\startdata
{\bf Aquarius dIrr (DDO 210)}     & 2013-06-27 & 1400s$\times$24 & & 2023-06-26 & 1230s$\times$12  & & \nodata & \nodata \\ 
{\bf Cetus dSph}    & 2006-08-30 & 1135s$\times$26 & & 2011-08-23 & 1267s$\times$\phn8     & & 2020-08-16    & 1234s$\times$\phn8 \\ 
{\bf Leo T}    & 2013-01-11 & 1050s$\times$24 & & 2015-11-19 & \phn950s$\times$24     & & 2024-01-21    & 1234s$\times$12 \\ 
{\bf Pegasus dIrr}    & 2015-07-23 & 1080s$\times$29 & & 2023-07-25 & 1230s$\times$20     & & \nodata    & \nodata \\ 
{\bf Pisces I (LGS~3)}    & 2005-09-14 & 1147s$\times$12 & & 2015-07-19 & 1189s$\times$24     & & 2023-09-12    & 1198s$\times$12 \\ 
{\bf Tucana dSph}    & 2006-04-27 & \phn957s$\times$32 & & 2011-05-12 & 1402s$\times$\phn8     & & 2020-05-10    & 1198s$\times$\phn8 \\
\enddata
\tablenotetext{a}{Exposure times in (seconds) $\times$ (number of exposures) format. 
                  Here we list the average across all exposures, 
                  but the actual individual exposure times vary by 
                  only a few percent in duration.}
\end{deluxetable*}

The sample of dwarf galaxies explored in this work is a subset of galaxies observed through HST GO-15911 and GO-17174. These proposals aimed to investigate the dynamics of intermediate mass dwarfs ($M_* \approx 10^5-10^7 \, \rm M_{\odot}$) using PM measurements. Further goals included connecting derived orbital histories with measured star formation histories (SFHs) to link the environment these dwarfs have spent their lifetimes in with their stellar assembly. Table \ref{tab:Props} summarizes the observed distances ($\rm D_{helio}, \, D_{M31}$) and line-of-sight velocities ($\rm V_{los, helio}$) from the literature that we adopt in this work. 

This work uses data from the \textit{Hubble Space Telescope} (HST) to derive PMs for the six dwarfs in our sample, which are discussed in Section \ref{subsec:pms} and listed in Table \ref{tab:Props}. Additionally, we have obtained Green Bank Telescope (GBT) observations of the HI distribution in Cetus. The latter improves upon previous observations from \citet{Putman_2021}, and will be further discussed in Section \ref{subsec:cetus} when we relate our orbital results to Cetus's SFH and gas content. 

\subsection{Hubble Space Telescope Observations}

All images used for the PM measurements of these dwarf galaxies were obtained with \textit{HST} Advanced Camera for Surveys/Wide Field Channel (ACS/WFC) using the F814W or F606W filters. Details of the epochs of observations and individual exposures are in Table~\ref{tab:epochs}. These data were taken as part of several observing programs across the past 20 years, including GO-10505, GO-12273, GO-12914, GO-12925, GO-13738, GO-13768, GO-13770, GO-15911, \& GO-17174. All data were downloaded from the Mikulski Archive for Space Telescopes (MAST).

\subsection{Green Bank Telescopes observations}
\label{subsec:gbt}

Of our dwarf sample, only Cetus and Tucana do not have well-established literature HI masses \citep[][]{Putman_2021}. Measurements of HI masses will be discussed in Section \ref{sec:discussion}, as they provide insight into the environment these dwarfs have survived in. To potentially detect HI in these dwarfs, we observed them with the GBT concurrently with this analysis. Tucana is a southern-sky object and not visible from the GBT, so only Cetus was observed.

This observation occurred during the 2024B semester under project ID GBT24B-319. The source was observed in 15 sets of 600-second ON/OFF pairs (300 s ON, 300 s OFF) for 75 minutes on-source.  The OFF position was chosen to avoid galactic HI emission at a similar velocity to Cetus. The L-band receiver was coupled with the Versatile GBT Astronomical Spectrometer (VEGAS) operating in Mode 10, providing a bandwidth of 23.44 MHz and a spectral resolution of 0.7 kHz (0.15 km/s). The spectral setup was centered at 1.42041 GHz (the HI 21 cm line) and then redshifted to account for the velocity of Cetus. The L-band receiver provides a beam size of approximately 9 arcminutes. 
These observations produced a non-detection (i.e., there is no detectable HI reservoir in Cetus), assuming a 15 km/s line width and velocity resolution, with a $5\sigma$ upper limit of 60.8 mJ km/s, or an HI mass of $M_{HI,lim} = 8\times10^{4}\, \rm M{_\odot}$ at the distance of Cetus (748~kpc).

\begin{deluxetable*}{c|c|c|c|c|c|c|c}[h]
\tablecaption{Dwarf Galaxy sample and properties}
\tablehead{\colhead{Dwarf} & \colhead{$ D_{\rm helio}$ (kpc)} & \colhead{$ D_{ \rm M31}$ (kpc)} & \colhead{$ V_{\rm los,helio}$ (km s$^{-1}$)} & \colhead{$\mu^{*}_{\alpha}$ (mas yr$^{-1}$)} & \colhead{$\mu_{\delta}$ (mas yr$^{-1}$)} & \colhead{$V_{\rm trans, \alpha}$ (km s$^{-1}$) } & \colhead{$V_{\rm trans, \delta}$ (km s$^{-1}$)} }
\startdata
Aquarius     & 1030$\pm$60  & 1137$\pm$42    & $-$140$\pm$1      & -0.0179$\pm$0.0123      & -0.0477$\pm$0.0123     & -87.4$\pm$60.1      & -232.9$\pm$60.1           \\ 
Cetus      & \phn748$\pm$32   &754$\pm$12    & $-$87$\pm$2     & \phn0.0618$\pm$0.0106    & -0.0730$\pm$ 0.0100      & 219.1$\pm$ 37.6      & -258.8$\pm$35.5           \\ 
Leo T      & \phn409$\pm$29     & 979$\pm$18 & 38$\pm$2      & -0.0259$\pm$0.0148      & -0.1286$\pm$0.0148      & -50.2$\pm$28.7      & -249.3$\pm$28.7           \\ 
Pegasus      & \phn887$\pm$21  & 457$\pm$10    & $-$183$\pm$1      & \phn0.0268$\pm$0.0098      & \phn0.0085$\pm$0.0098      & 112.7$\pm$41.2      & 35.7$\pm$41.2           \\ 
Pisces     & \phn604$\pm$14     & 293$\pm$6 & $-$287$\pm$1      & \phn0.0230$\pm$0.0100      & \phn0.0050$\pm$0.0099      & 65.8$\pm$28.6      & 14.3$\pm$28.3         \\ 
Tucana      & \phn855$\pm$36  & 1323$\pm$28    & 194$\pm$4     & \phn0.0135$\pm$0.0074      & -0.0224$\pm$0.0074     & 54.7$\pm$30.0     & -90.1$\pm$30.0           \\ \hline
\enddata
\tablecomments{$\mu^{*}_{\alpha}$ = $\mu_{\alpha} \times cos(\delta),$ where $\mu_\alpha$ is the time derivative of the RA in units of angle (not hour, minute, second units). These are on-sky values and are not corrected for solar reflex motion. $V_{\rm trans, \alpha}$ and $V_{\rm trans, \delta}$ refer to the transverse velocities computed using the proper motions and distances provided. Uncertainties on $V_{\rm trans, \alpha}$ and $V_{\rm trans, \delta}$ do not include the propagation of distance uncertainties. References for the literature heliocentric distances, column 2: Aquarius \citep[][]{Ordo_2016}, Cetus \& Tucana \citep[][]{Dambis_2013}, Leo~T \citep[][]{Clementini_2012}, Pegasus \& Pisces \citep[][]{Savino_2025}. References for the literature LOS velocities, column 4: Aquarius \citep[][]{Springob_2005}, Cetus \citep[][]{Kashibadze_2018}, Leo~T \citep[][]{Haynes_2018}, Pegasus \& Pisces \citep[][]{Huchtmeier_2003}, Tucana \citep[][]{Fraternali_2009}. }
\label{tab:Props}
\end{deluxetable*}

\subsection{Proper Motion Analysis}
\label{subsec:pms}

To measure the PMs of our dwarf sample, we closely follow the methods used extensively in our previous work \citep{Sohn_2012, Sohn_2013, Sohn_2017, Sohn_2020, Bennet_2024}. We will summarize our methodology here, but invite interested readers to read these other works for more details. 

We started by downloading the fully pipeline-processed image for each exposure. Specifically, we downloaded
the {\tt \_flc.fits} images, which not only include all standard CCD reduction steps (bias and dark subtraction, flat-fielding, etc.), but also have been processed for the imperfect charge transfer efficiency (CTE) using the correction algorithms of \citet{Anderson_2010}.
We determine a position and flux for each point source from the {\tt \_flc.fits} images using the FORTRAN code {\tt hst1pass} \citep{Anderson_2022}. We applied corrections to the positions using the methodology from \citep[][]{Anderson_2006}, with updated ACS/WFC geometric distortions based on \citet{KP_2015}; these were further extended to include time-dependent distortion variations beyond 2020 (V. Kozhurina-Platais, private communication). For each dwarf galaxy, we create high-resolution stacked images with pixel scales of $0.025$ mas using images from the reference epoch. We use the epoch with the most exposures as the reference epoch across our sample (see Table \ref{tab:epochs} for exposure counts for each epoch). 

We identify point sources associated with the dwarf galaxies via color-magnitude diagrams (CMDs) constructed from multi-band photometry. To construct CMDs, we use images obtained during the first-epoch observations for all targets. These first epoch data include images from other bands (F475W or F606W, depending on the dwarf galaxy) that allow us to create these CMDs. Background galaxies were identified first through an objective selection based on the quality-of-fit parameter output from the {\tt hst1pass} code, then confirmed via visual inspection of each source. 

We then construct an empirical ``template'' for each point source and background galaxy by supersampling the objects extracted from the high-resolution stack of images. Supersampling uses the fact that the dithers between \textit{HST} exposures are a non-integer number of pixels; thus, stacked images are super-sampled by a factor of two relative to the native ACS/ WFC pixel scale for better spatial resolution. 
Each template takes into account the point-spread function (PSF), the morphology of background galaxies, and the pixel binning \citep[][]{Sohn_2013}. These templates are fitted on each individual {\tt \_flc.fits} image for measuring the position of each point source or galaxy. We fit the templates directly onto the individual images for the reference epoch since the templates were constructed {\it from} that epoch. For the other non-reference epoch(s), we employ $7\times7$ pixel convolution kernels when fitting templates to allow for differences in PSF between the different epochs. These convolution kernels were derived by comparing PSFs of numerous bright and isolated point sources associated with the target dwarf galaxies between the reference and other epochs \citep[][]{Sohn_2013}. At this stage, we are left with template-fitted positions of all point sources and background galaxies in all individual images in each epoch.

We define the reference frame for each dwarf galaxy in our sample (the frame with respect to which motions are measured) by averaging the template-based positions of point sources from repeated exposures of the reference epoch. The positions of point sources in each exposure of the other epochs were then used to transform the template-measured positions of the galaxies into the reference frames. Subsequently, we measured the positional difference for each background galaxy between the reference and other epochs, relative to the stars associated with the dwarfs. 

Local corrections were applied to remove any residual CTE and remaining geometric distortion systematics by making measurements only relative to stars with similar brightness in the local vicinity on the image.

Once the reflex displacements of background galaxies are measured for each non-reference epoch, we calculate the final PMs. For those dwarfs with only two epochs of observations, i.e., Aquarius and Pegasus, the PM calculations were carried out straightforwardly following those in \citet{Sohn_2020}. In summary, we take the error-weighted average of all displacements of background galaxies with respect to dwarf member stars for each individual first-epoch exposure to obtain an independent PM estimate for each exposure. The PM was obtained from the error-weighted mean weighted by the inverse variance of individual PM exposure estimates. Similarly, associated uncertainties in the mean were obtained by summing the inverse variances to get the inverse variance of the mean, also from the individual exposure PM estimates.

For the rest of the sample, where we have three epochs, we first calculate the average position of each background galaxy in each epoch. This was done by directly adopting the averaged positions for the reference epoch, and by converting the local-corrected displacements into positions for the additional epochs. 
Because we have multiple exposures and measurements in each epoch, the mean position of each background galaxy in each epoch has a known associated uncertainty. We fit straight lines through the background galaxies X and Y positions as a function of $\Delta T$, being the time since the first epoch in units of years, using chi-square minimization of the residuals weighted with the position measurement uncertainties \footnote{Using only the initial and final epoch for the dwarfs where three epoch analysis is possible, to replicate the two epoch analysis technique, does yield consistent results.}. The resulting slope of this fitting process is the PM of each background galaxy relative to the member stars. The PM and associated uncertainty of the dwarf galaxies were then obtained by the error-weighted mean of these individual background galaxy PM estimates (see Table \ref{tab:Props}).

\begin{deluxetable*}{lcccccc}
\tablecaption{Local Group Dwarfs Position and Velocity Vectors \label{tab:Pos_vel}}
\tablewidth{0pt}
\tablehead{
\colhead{Dwarf}  &  \colhead{$x$} & \colhead{$y$} & \colhead{$z$} &  \colhead{$v_x$} & \colhead{$v_y$} & \colhead{$v_z$} \\
\multicolumn1c{} & \colhead{(kpc)} &  \colhead{(kpc)} & \colhead{(kpc)} &  \multicolumn1c{(km $s^{-1}$)} & \multicolumn1c{(km $s^{-1}$)}  & \multicolumn1c{(km $s^{-1}$)} 
}
\startdata
Aquarius & 719 $\pm$ 42 & 493 $\pm$ 29 & -538 $\pm$ 31 & 46 $\pm$ 43 & -32 $\pm$ 55 & 67.3 $\pm$ 51 \\
Cetus & -54 $\pm$ 2 & 216 $\pm$ 9 & -715 $\pm$ 31 & -40 $\pm$ 37 & -100 $\pm$ 38 & -2.4 $\pm$ 11 \\
Leo T & -250 $\pm$ 17 & -169 $\pm$ 12 & 283 $\pm$ 20 & 34 $\pm$ 23 & 0 $\pm$ 31 & -65.9 $\pm$ 22 \\
Pegasus & -63 $\pm$ 1 & 641 $\pm$ 15 & -611 $\pm$ 14 & -91 $\pm$ 41 & 92 $\pm$ 29 & 122.0 $\pm$ 30 \\
Pisces & -282 $\pm$ 6 & 366 $\pm$ 8 & -395 $\pm$ 9 & 84 $\pm$ 26 & 43 $\pm$ 23 & 209.5 $\pm$ 22 \\
Tucana & 452 $\pm$ 19 & -349 $\pm$ 15 & -630 $\pm$ 26 & 71 $\pm$ 25 & 73 $\pm$ 28 & -117.5 $\pm$ 21 \\
\enddata
\tablecomments{Three-dimensional position and velocity vectors derived, in Galactocentric coordinates, from the galaxy positions, distances, LOS velocities, and PM measurements as reported in Table~\ref{tab:Props}. The Galactocentric distance of the Sun and the circular velocity of the local
standard of rest (LSR) are adopted from \citet{mcmillan11}. Solar peculiar velocities with respect to the LSR are taken from \citet{Schonrich_2010}. }
\end{deluxetable*}

\section{Orbits of the LG Dwarfs} \label{sec:Orbits}
\subsection{Galactocentric Coordinates}
\begin{figure*}[ht]
\centering
\includegraphics[width=\textwidth]{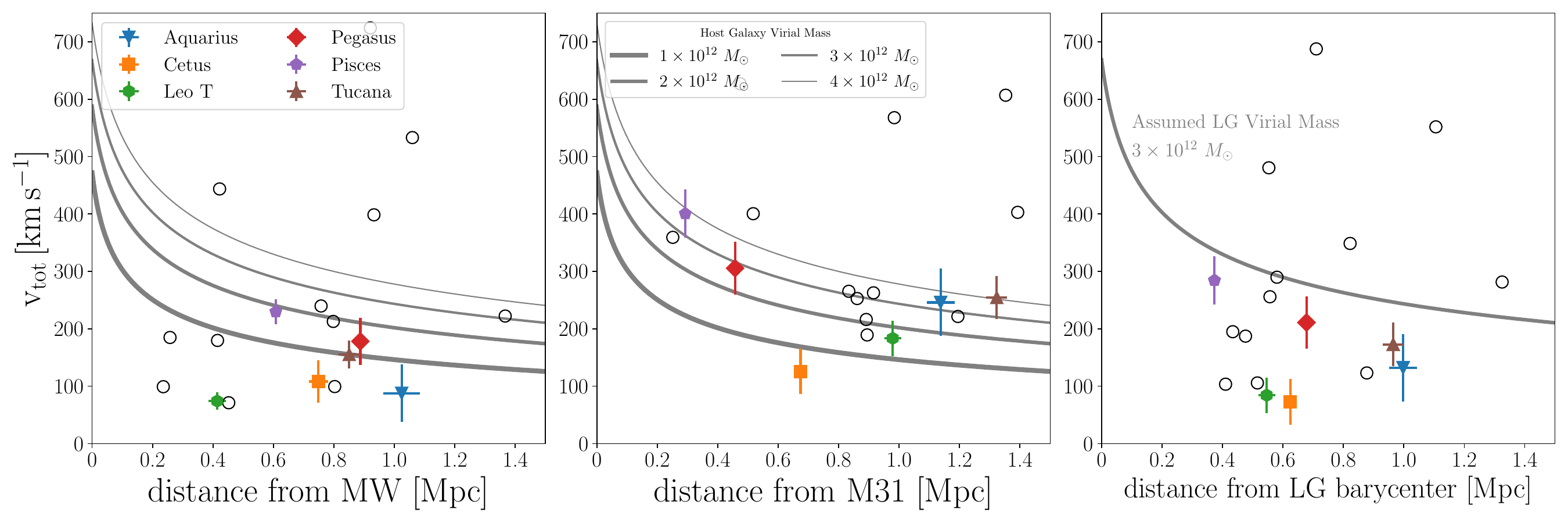}
    \caption{3D position with respect to the MW (left panel) and M31 (middle panel) relative to 3D velocity for each dwarf as reported in Table~\ref{tab:Pos_vel}. Escape velocity curves are also shown as gray lines of varying thickness for host galaxy halos with masses between $1-4\times10^{12} \, \rm M_{\odot}$ assuming NFW dark matter halos. Open circles show similar results (without error bars) for the LG periphery dwarfs published in \citetalias{Bennet_2024}. Four and six dwarfs are bound to the MW and M31, respectively, for the masses adopted in this work, which are $1\times10^{12}\, \rm M_{\odot}$ for the MW and $2\times10^{12}\, \rm M_{\odot}$ for M31. The right panel shows the dwarf data with respect to the LG barycenter, assuming a 1:2 mass ratio for the MW and M31, respectively (i.e., $3\times10^{12}\, \rm M_{\odot}$ together). In our assumed LG virial mass model, all six dwarfs are bound to the LG. }
    \label{fig:escape}
\end{figure*}

Using the LOS velocity and distance measurements compiled in Table~\ref{tab:Props} combined with the PMs measured in this work (see also Table~\ref{tab:Props}), 6D phase space information is derived for each dwarf galaxy in Galactocentric Cartesian coordinates and reported in Table~\ref{tab:Pos_vel}. Uncertainties on the position and velocity components reflect the standard deviation measured from the propagation of 1$\sigma$ uncertainties on LOS velocity, distance, and PM. The listed uncertainties are highly correlated because the error ellipses are not aligned with Galactocentric coordinates
and have large axial ratios: position uncertainties are much smaller in the on-sky directions than in the line-of-sight direction,
while velocity uncertainties are much larger in the on-sky directions than in the line-of-sight direction.


Figure~\ref{fig:escape} shows the magnitude of the 3D velocity for each dwarf relative to the distance from the MW (middle panel) and M31 (right panel) compared to escape velocity curves for host galaxy halos of masses between $1-4\times10^{12}\, \rm M_{\odot}$. This mass range captures the literature mass estimates for both the MW and M31 \citep[see reviews for MW and M31, respectively;][]{wang20, bhattacharya23}. The relative distances and velocities are computed assuming the distance and PM values for M31 listed in Section \ref{subsec:orbit_methods}.

Escape velocity curves were computed assuming NFW halo profiles. This work uses a virial mass of $1\times10^{12}\, \rm M_{\odot}$ for the MW and $2\times10^{12}\, \rm M_{\odot}$ for M31. However, only four of the six dwarfs are bound to the MW assuming $\rm M_{vir}=1\times10^{12}\, \rm M_{\odot}$ \citep{Patel_2018}.
For M31 \citep[$\rm M_{vir} =2\times10^{12}\, \rm M_{\odot}$;][e.g.,]{Fardal_2013, VillanuevaDomingo22}, only two of six dwarfs are bound to M31. While not all dwarfs are bound to either the MW or M31, they are bound to the joint potential of the MW+M31.  

If we instead compute the magnitude of 3D position and velocity relative to the barycenter of the LG, as shown in the right-most panel of Figure~\ref{fig:escape}, assuming the total LG mass is the sum of the MW and M31's masses ($3\times10^{12}\, \rm M_{\odot}$), all six dwarfs have velocities such that $\rm v_{tot} < v_{esc}(d)$, implying that all dwarfs are bound to the LG at present.

Open circles in all panels of Figure~\ref{fig:escape} represent data for other dwarf galaxies at the periphery of the LG from \citetalias{Bennet_2024}. There is a significant intersection between these previously published dwarfs and the sample presented here, but no dwarfs in our sample exhibit very high velocities ($\rm > 400\, \rm km\,s^{-1}$) as in \citetalias{Bennet_2024}.

\subsection{Orbit Methodology}
\label{subsec:orbit_methods}
\begin{table}[]

    \centering
    \caption{Dwarf Galaxy Potential Parameters}
    \begin{tabular}{c|c|c}\hline \hline
    Galaxy & $r_P$[kpc] & $M_{halo}\, [10^{10}\, \rm M_{\odot}]$   \\ \hline \hline
     Aquarius    &  5.34 & 0.6 \\ 
     Cetus    & 0.0001  & 0.8\\ 
     Leo T & 2.46 & 0.5\\ 
     Pegasus & 0.0001 & 1.5 \\ 
     Pisces &  0.0001 & 0.5 \\ 
     Tucana & 0.0001 & 0.5 \\  \hline
    \end{tabular}
    \tablecomments{Halo masses estimated from the \citet{moster13} abundance matching relation and Plummer scale lengths ($r_P$) derived by fitting the dynamical mass for each galaxy to a Plummer profile. Dynamical masses are computed with the \citet{wolf10} mass estimator. Stellar masses, half-light radii, and velocity dispersions are taken from \citet{McConnachie_2012}. A Plummer scale length of 0.0001 kpc represents galaxies modeled as point masses.}
    \label{tab:dwarf_params}
\end{table}

To reconstruct orbital histories for the six LG dwarfs in our sample, we follow the methodology of \citetalias{Bennet_2024}. In brief, the 6D phase space coordinates reported in Table~\ref{tab:Pos_vel} are used to initialize backward numerical orbit integration for 6 Gyr. We also compute orbits 6 Gyr into the future. Our chosen integration period of 6 Gyr corresponds to the time by which the MW and M31 had assembled about 80\% of their mass \citep{santistevan20}. 
For each galaxy, we compute a 5-body orbit accounting for the gravitational potential of the MW, M31, LMC\footnote{Note that we do not correct for the perturbation introduced by the LMC, which offsets the outer MW halo from the inner MW halo \citep{GaravitoCamargo_2019}. As discussed in \ref{sec:Orbits}, none of our six dwarfs closely approach the MW or the LMC, so this effect is likely negligible for the purposes of this work.}, M33, and each dwarf galaxy of interest. For the four massive LG galaxies, we use identical masses and parameters as those in \citetalias{Bennet_2024}, which are originally adopted from \citet{patel17a}. Each galaxy is represented as an extended mass distribution. The MW and M31 are represented as three-component potentials with an NFW halo, Miyamoto-Nagai disk, and a Hernquist bulge. Dark matter halos are adiabatically contracted using the \texttt{CONTRA} code \citep{contra}. Both the LMC and M33 are modeled as single-component Plummer spheres. Assumed total masses for each galaxy are as follows: $(2, 1, 0.25, 0.18) \times  10^{12}\, \rm M_{\odot}$ for M31, the MW, M33, and the LMC, respectively. 

All galaxies are allowed to move in response to each other's gravitational forces, with the exception of the six lower mass dwarf galaxies, which do not exert gravitational forces on the other four galaxies. All dwarfs except Aquarius and Leo T are modeled as point masses. These two galaxies are each modeled as Plummer spheres with the scale radius and halo mass listed in Table~\ref{tab:dwarf_params}. Dynamical friction owing to the MW and M31 is included. All galaxy potentials are rigid, i.e., the potentials and masses remain fixed as the centers of mass are advanced at each integration time step.

We use M31 and M33 distances from \citet[776.2 kpc and 859 kpc, respectively;][]{Savino_2022}, M31's PM from \citet{Salomon_2021}, and M33's PM from \citet{Brunthaler_2005}. 6D phase space properties for the LMC are identical to those in \citet{patel17a}. We encourage interested readers to see \citetalias{Bennet_2024} and \citet{patel17a} for model parameters and details relevant to the orbital integration scheme.

Orbits initialized with the mean 6D phase space coordinates listed in Table \ref{tab:Pos_vel} will be referred to as direct orbital histories, as in past work \citep{Sohn_2020, Patel_2020, Bennet_2024}. Uncertainties on orbital histories will be discussed in Section \ref{subsec:orbit_errors}.

\subsection{Orbital Histories and Analysis}
\label{subsec:orbits}

\begin{figure*}[h]
\centering
    \includegraphics[scale=0.6]{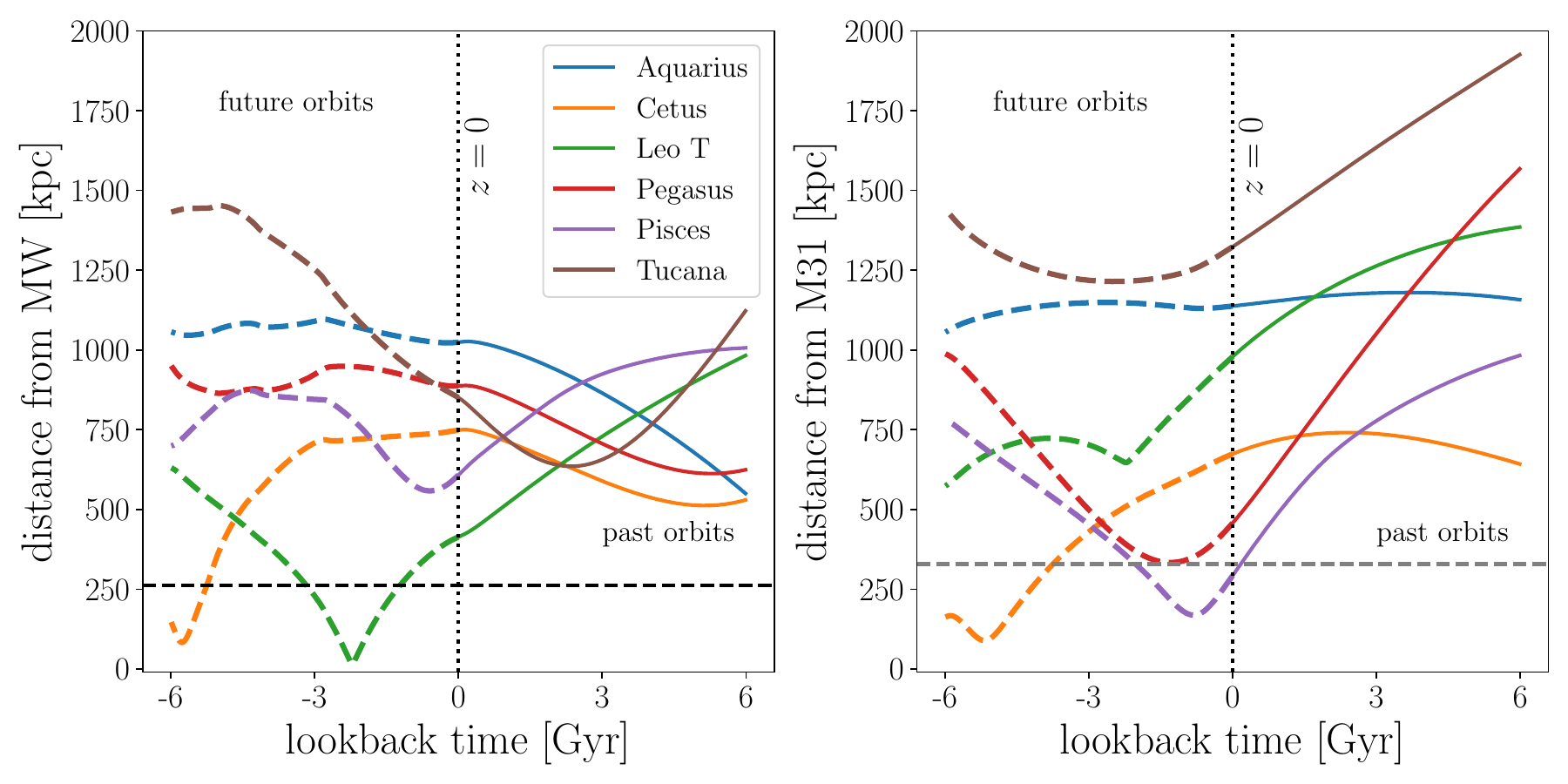}
    \includegraphics[scale=0.6]{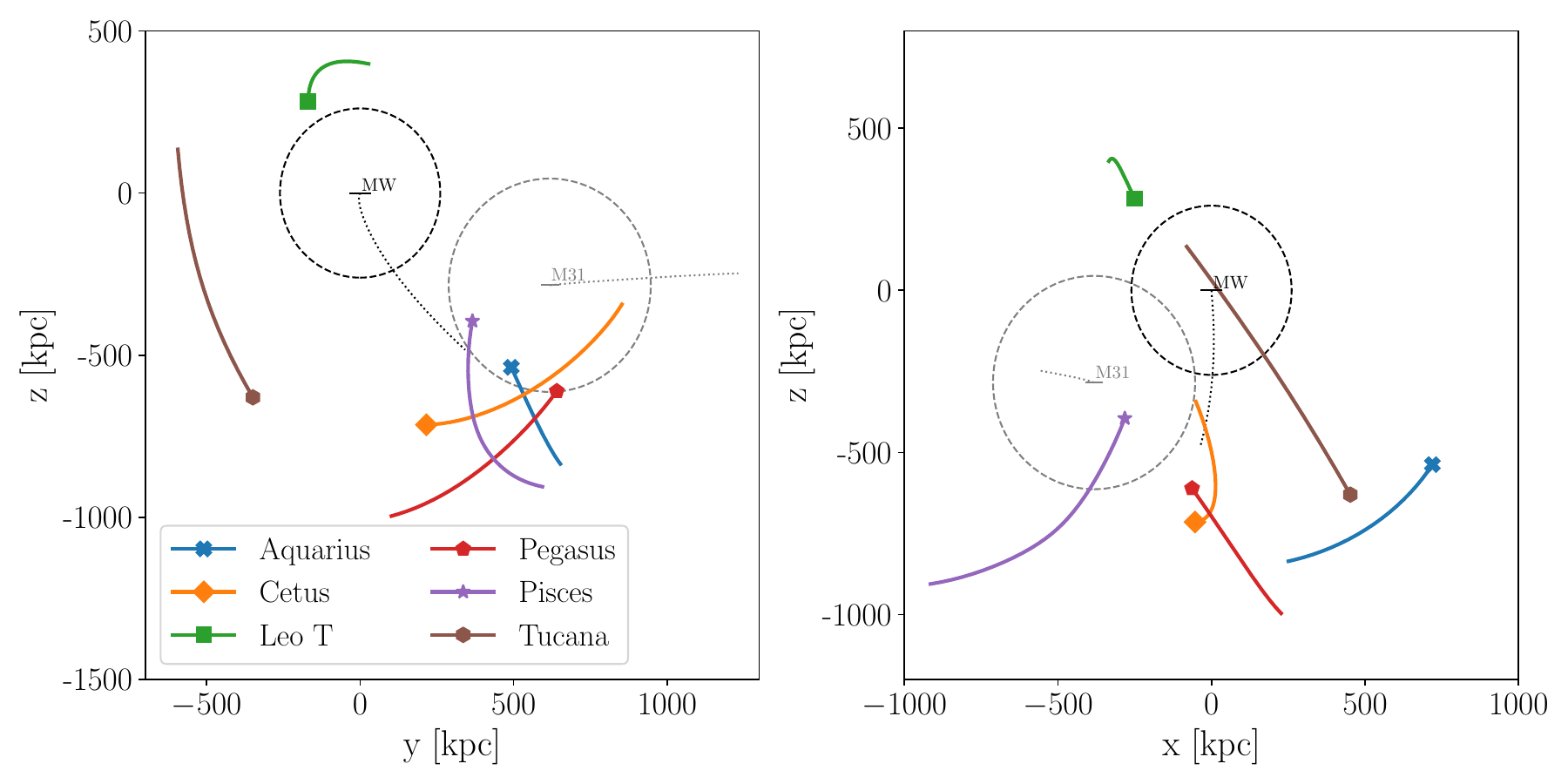}
    \caption{\textbf{Top:} Direct orbital histories for all six galaxies in our sample spanning the range of 6 Gyr into the future to 6 Gyr in the past. The present day is denoted with a dotted vertical line. The right of this vertical line represents past orbits, while the left illustrates future orbits. Orbits are computed in a joint potential accounting for the MW, M31, M33, and the LMC following the methods of \citetalias{Bennet_2024}. The left panel shows orbits relative to the MW, while the right panel shows orbits with respect to M31. Orbits are initialized with the 6D phase space coordinates reported in Table \ref{tab:Props}. The black and gray dashed lines indicate the virial radii of the MW and M31, respectively. Based on the direct orbital histories, none of the dwarfs have passed within the virial radius of the MW (261 kpc) or M31 (329 kpc) in the past 6 Gyr, however, Leo T and Cetus may make a close passage with the MW in the future, and Pisces may be categorized similarly with respect to M31. \textbf{Bottom:} Spatial projection of the direct orbital histories provided in Figure \ref{fig:orbits} in Galactocentric coordinates. The virial extent of the MW (black) and M31's (gray) halos is indicated by dashed circles. Similarly, the dotted lines show the orbits of the MW and M31 over the last 6 Gyr. For all dwarfs, markers indicate their present-day positions, and lines indicate their trajectories backward. All dwarfs except Pisces have remained outside the virial radii of both the MW and M31 over the past 6 Gyr. }
    \label{fig:orbits}
\end{figure*}

Direct orbital histories for all six dwarfs are illustrated in the top panels of Figure~\ref{fig:orbits}. The left panel shows orbits relative to the MW, while the right panel shows orbits with respect to M31. Orbits are integrated into the past (solid lines) and future (dashed lines) for 6 Gyr, separated by the dotted vertical line. The dashed horizontal lines in each panel illustrate the virial radii of the MW ($\rm R_{vir}=261$ kpc) and M31 ($\rm R_{vir}=329$ kpc). 

With the exception of Pisces, which only recently entered the virial radius of M31, Figure~\ref{fig:orbits} shows none of the dwarfs have passed within the virial radius of the MW or M31 within the last 6 Gyr (i.e., backsplash galaxies), but some may interact with the MW (Cetus, Leo T) or M31 (Pisces) in the future. In Section \ref{subsec:orbit_errors}, we will see that statistical ensembles on the orbital properties allow for small backsplash probabilities.

The bottom panels of Figure \ref{fig:orbits} show the direct orbital histories as cross sections. The virial radii of the MW and M31 are now shown as black and gray dashed circles. Orbits are only illustrated for the last 6 Gyr in this view. Markers represent the dwarfs' location today, and the lines trace backward in time. Orbits of the MW and M31 are also indicated as dotted lines, as they are free to move in response to each other and the LMC and M33's gravitational influence. While we will not discuss past interactions with the LMC or M33 further in this paper, we refer readers to a companion paper on this topic, highlighting Pisces in particular (Patel, Bennet, et al., in prep.). 

Figure \ref{fig:orbits} shows that Pisces is the only galaxy within the virial radius of either massive galaxy (M31 in this case) at the present day, and therefore the only M31 satellite\footnote{We will refer to ``satellites" as galaxies that are currently within the virial radius of a more massive halo for consistency with observational surveys. ``Bound satellites" will refer to satellites currently within the virial radius of a more massive halo that also have a 3D velocity less than the equivalent escape velocity at the same relative distance to the host.} in the dwarf sample. While the right panel of Figure \ref{fig:orbits} shows that Tucana crosses through the MW's halo, the two cross sections combined indicate that this is actually a projection effect, and Tucana is actually quite far from the MW ($\sim$800 kpc) and M31 ($\sim$1300 kpc). Both cross sections show that Leo T resides near the boundary of the MW's halo, hinting at the future interaction previously illustrated in the top left panel of Figure \ref{fig:orbits}.

\subsection{Orbital History Uncertainties}
\label{subsec:orbit_errors}

To understand how statistically common the orbital histories illustrated in Figure ~\ref{fig:orbits} are and to identify any backsplash galaxies, we must account for measurement uncertainties on the LOS velocity, distance, and PM in the numerical orbit integrations\footnote{In this work, we do not test alternative MW and M31 masses and their effect on the orbits of the six dwarfs, as in \citet{Patel_2020}. However, as noted in our previous work, the assumed masses of the MW and M31 do significantly contribute to the orbital uncertainties. For consistency with \citetalias{Bennet_2024}, we only test a fixed set of masses in this work and leave further investigation of how different MW and M31 masses affect the result presented here to future work}. 

\begin{figure*}
\centering
\includegraphics[width=\textwidth]{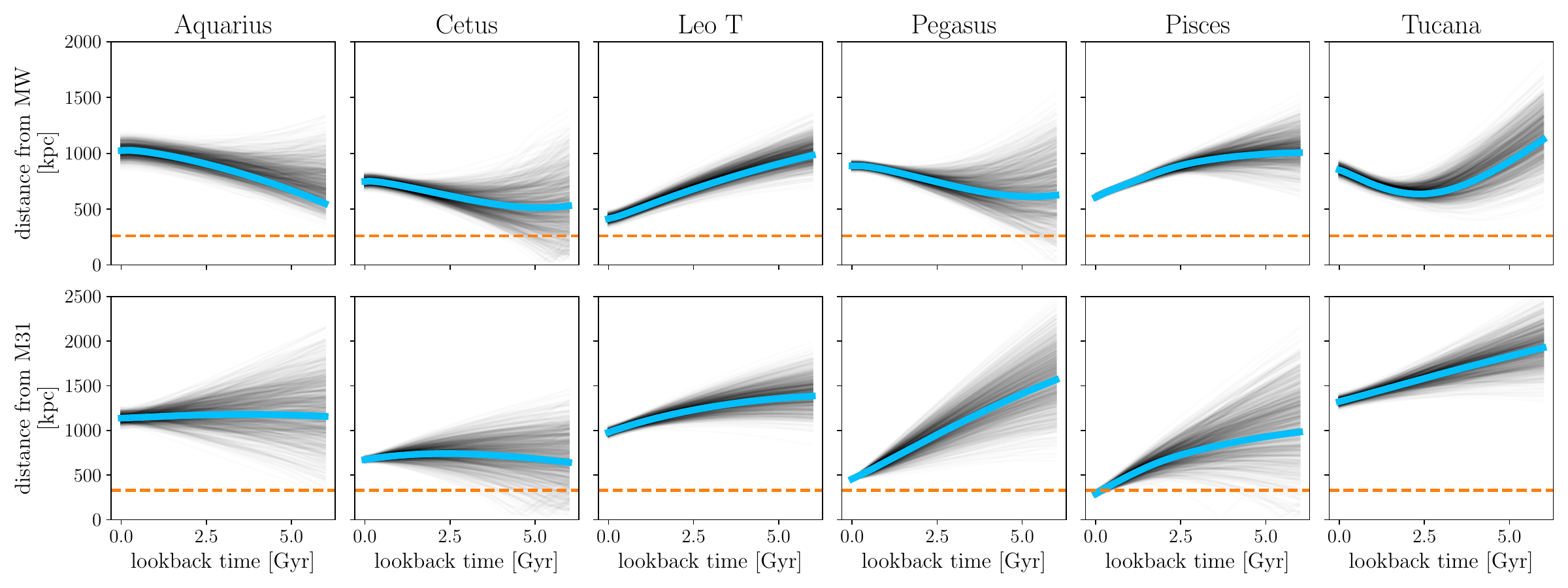}
    \caption{Orbital uncertainties for all dwarfs with respect to the MW (top panel) and M31 (bottom panel) are shown in gray, while the blue lines indicate direct orbital histories from Figure \ref{fig:orbits}. The dashed orange lines indicate the virial radii of the MW and M31. Tables \ref{tab:orbit_params_mw} and \ref{tab:orbit_params_m31} quantitatively summarize the range of plausible orbital histories for all dwarfs. Leo T, Tucana, and Pisces are most statistically likely to be on their first passage through the MW's halo. While Tucana has completed a pericenter around the MW, Leo T and Pisces have yet to reach their pericenter about the MW. Aquarius, Cetus, and Pegasus are more statistically likely to have reached both pericenter and apocenter about the MW. Still, these orbits do not typically bring the dwarfs closer than 500 kpc from the MW. Leo T, Pegasus, and Tucana are on first infall into M31's halo. While most Pisces' orbits indicate first infall into M31, a past interaction is possible for a small fraction of orbits. For Cetus and Aquarius, passages around M31 are statistically common. Cetus passes significantly closer to M31 than Aquarius does. Aquarius does not reach within 500 kpc of M31, while Cetus does.} 
    \label{fig:errors}
\end{figure*}

As in \citetalias{Bennet_2024}, we run 1,000 orbits for each of the six dwarfs, initializing each orbit with a 6D phase space vector drawn from the joint measurement uncertainties in a Monte Carlo fashion. These joint measurement uncertainties also include uncertainties for M31, the LMC, and M33. The resulting range of orbital histories for each galaxy relative to the MW and M31 are shown in the top and bottom rows of Figure \ref{fig:errors}, respectively. The blue lines show the same direct orbital histories from Figure \ref{fig:orbits}. Summary statistics for orbital parameters derived from the set of 1,000 orbits per dwarf are provided in Tables \ref{tab:orbit_params_mw} and \ref{tab:orbit_params_m31}. 

From here on, \emph{first passage} refers to galaxies that completed a pericenter recently about a more massive galaxy and are now moving away from the host galaxy \citep[e.g., Leo~I][]{BoylanKolchin_2013,Sohn_2013}, while \emph{first infall} refers to galaxies where a pericenter has not yet occurred, but the galaxy is still moving towards a more massive galaxy. \emph{Backsplash} is defined as one (or more) pericenters within the virial radius of the MW or M31 in the last 6 Gyr.

The summary statistics for dwarf orbital parameters relative to the MW (Table \ref{tab:orbit_params_mw}) show that Leo T, and Pisces are likely on first infall, as only a tiny fraction ($\leq2$ \%) of orbits have a pericentric passage around the MW and were $\sim$1000 kpc from the MW at 6 Gyr ago, on average. 

On the other hand, Aquarius, Cetus, Pegasus, and Tucana have a significant fraction of orbits where at least a pericenter, if not also an apocenter, exists in the last 6 Gyr. Tucana is the only dwarf in the sample on its first passage around the MW as it originated far from the MW and completed a pericenter about 2.3 Gyr ago.

Aquarius, Cetus, and Pegasus have similar orbital histories, such that many of the 1,000 orbits indicate the dwarfs have completed both a pericenter and apocenter about the MW, but typically at large distances ($\gtrsim$ 500 kpc).

Furthermore, Cetus and Pegasus even have some orbits where the distance at pericenter was within the virial radius of the MW. Pegasus only reaches within the virial radius of the MW in 0.8\% of orbits. Cetus exhibits a pericenter within the MW's virial radius for 5.6\% of all orbits (see Table \ref{tab:splash} and Section \ref{subsec:cetus}).  

For certain dwarfs, such as Aquarius, extrapolating from the orbits shown in the top row of Figure \ref{fig:errors}, it is possible that it interacted with the MW beyond 6 Gyr ago, but again, this is only speculation as we do not advise integrating beyond the 6 Gyr time period \citep[see][]{Bennet_2024, Patel_2020, santistevan20}. 

Turning to the summary statistics computed with respect to M31 (Table \ref{tab:orbit_params_m31}), Pegasus, Leo T, and Tucana are on first infall, never having passed around M31 in the past 6 Gyr. Leo T has a non-zero fraction of apocenters relative to M31, but all the distances reached at apocenter are at $\gtrsim$ 1000 kpc from M31. 

For Pisces, only $\sim$2\% of orbits have a pericenter, but the distances at pericenter are typically within the virial radius of M31. These distances range from $\sim$30-600 kpc, but only 1.3\% of all Pisces orbits have a pericenter strictly within M31's virial radius. The possibility of past interactions between Pisces and M33-M31 will be further discussed in a companion paper (Patel, Bennet, et al., in prep.).

Finally, Aquarius and Cetus have the most significant fraction of orbits where pericenters and apocenters are common about M31. However, the primary difference between the orbital parameters of these dwarfs is that Cetus has a non-zero fraction of orbits (4.4\%) where a pericenter occurs within M31's virial radius. In contrast, all Aquarius orbits where a pericenter exists occur at distances greater than approximately three times M31's virial radius. We conclude Cetus may be a backsplash galaxy of M31 and will discuss further details in Section \ref{subsec:cetus}.

\begin{deluxetable*}{lcccccc}[h]
\tablecaption{Orbital Parameters Relative to the Milky Way\label{tab:orbit_params_mw}}
\tablewidth{0pt}
\tablehead{
\colhead{Dwarf}  &  \colhead{$\rm f_{peri}$} & \colhead{$\rm t_{peri}$} & \colhead{$\rm r_{peri}$} &  \colhead{$\rm f_{apo}$} & \colhead{$\rm t_{apo}$} & \colhead{$\rm r_{apo}$} \\
\multicolumn1c{} & \colhead{(\%)} &  \colhead{(Gyr)} & \colhead{(kpc)} &  \multicolumn1c{(\%)} & \multicolumn1c{(Gyr)}  & \multicolumn1c{(kpc)} 
}
\startdata
Aquarius & 23 & $\cdots$ [3.01, 4.4, 5.49] & $\cdots$ [790, 905, 996] & 100 & 0.19 [0.16, 0.18, 0.22] & 1026 [971, 1025, 1083] \\
Cetus & 81 & 5.14 [3.16, 4.61, 5.62] & 512 [394, 557, 678] & 100 & 0.13 [0.11, 0.12, 0.14] & 750 [720, 749, 780] \\
Leo T & 0 & $\cdots$ [5.81, 5.81, 5.81] & $\cdots$ [932, 932, 932] & 1 & $\cdots$ [5.53, 5.69, 5.73] & $\cdots$ [887, 925, 932] \\
Pegasus & 78 & 5.28 [2.7, 4.11, 5.38] & 612 [562, 696, 807] & 96 & 0.14 [0.09, 0.12, 0.15] & 888 [869, 887, 908] \\
Pisces & 2 & $\cdots$ [3.99, 4.96, 5.69] & $\cdots$ [809, 865, 892] & 28 & $\cdots$ [2.89, 4.7, 5.66] & $\cdots$ [859, 925, 984] \\
Tucana & 100 & 2.35 [1.81, 2.2, 2.75] & 635 [595, 652, 697] & 0 & $\cdots$ [0.0, 0.0, 0.0] & $\cdots$ [0, 0, 0] \\
\enddata
\tablecomments{Orbital parameters calculated relative to the MW ($\rm R_{vir}=261$ kpc). $\rm f_{peri}$ and $\rm f_{apo}$ denote the fraction of 1000 orbits where a pericenter or apocenter exists in the last 6 Gyr. Other parameters are taken from the direct orbital histories shown in Figure \ref{fig:orbits}, while values in brackets quote the [15.9, 50, 84.1] percentiles calculated from the set of 1000 orbits computed for each galaxy. } 
\end{deluxetable*}

\begin{deluxetable*}{lcccccc}[h]
\tablecaption{Orbital Parameters Relative to Andromeda\label{tab:orbit_params_m31}}
\tablewidth{0pt}
\tablehead{
\colhead{Dwarf}  &  \colhead{$\rm f_{peri}$} & \colhead{$\rm t_{peri}$} & \colhead{$\rm r_{peri}$} &  \colhead{$\rm f_{apo}$} & \colhead{$\rm t_{apo}$} & \colhead{$\rm r_{apo}$} \\
\multicolumn1c{} & \colhead{(\%)} &  \colhead{(Gyr)} & \colhead{(kpc)} &  \multicolumn1c{(\%)} & \multicolumn1c{(Gyr)}  & \multicolumn1c{(kpc)} 
}
\startdata
Aquarius & 14 & $\cdots$ [0.46, 1.68, 4.66] & $\cdots$ [1004, 1100, 1155] & 26 & 3.65 [1.53, 3.38, 5.29] & 1180 [1137, 1194, 1305] \\
Cetus & 13 & $\cdots$ [4.61, 5.27, 5.75] & $\cdots$ [105, 638, 773] & 62 & 2.38 [0.91, 2.22, 4.12] & 741 [687, 733, 818] \\
Leo T & 0 & $\cdots$ [5.96, 5.96, 5.97] & $\cdots$ [1388, 1461, 1534] & 16 & $\cdots$ [4.29, 5.03, 5.64] & $\cdots$ [1141, 1191, 1259] \\
Pegasus & 0 & $\cdots$ [0.0, 0.0, 0.0] & $\cdots$ [0, 0, 0] & 0 & $\cdots$ [5.54, 5.9, 5.93] & $\cdots$ [767, 889, 916] \\
Pisces & 2 & $\cdots$ [4.93, 5.63, 5.86] & $\cdots$ [32, 95, 572] & 18 & $\cdots$ [2.26, 3.86, 5.22] & $\cdots$ [508, 609, 718] \\
Tucana & 0 & $\cdots$ [0.0, 0.0, 0.0] & $\cdots$ [0, 0, 0] & 2 & $\cdots$ [3.96, 4.18, 5.0] & $\cdots$ [1408, 1459, 1546] \\
\enddata
\tablecomments{Same as Table \ref{tab:orbit_params_mw} but with respect to M31 ($\rm R_{vir}=329$ kpc)}
\end{deluxetable*}

\begin{deluxetable*}{c|c|c|c}[ht]
\tablecaption{Probability of dwarfs being a backsplash galaxy in the last 6 Gyr \label{tab:splash}}
\tablewidth{0pt}
\tablehead{
\colhead{Dwarf}  &  \colhead{Probability of} &  \colhead{Probability of} & \colhead{Probability of}  \\
\colhead{Name}  &  \colhead{Backsplash from MW} & \colhead{Backsplash from M31} & \colhead{Backsplash Total} \\
\multicolumn1c{} & \colhead{This Work} & \colhead{This Work} &  \colhead{\cite{Buck_2019}}
}
\startdata
Aquarius & 0 & 0 & - \\
Cetus & 5.6 & 4.4 & 36 \\
Leo T &  0 & 0 & 51 \\
Pegasus & 0.8 & 0 & 79 \\
Pisces & 0 & 1.3 & - \\
Tucana & 0 & 0 & - \\ \hline
\enddata
\tablecomments{Probabilities of dwarf galaxies in our sample being backsplash galaxies compared to those computed using radial velocity and position in \cite{Buck_2019} where available.}
\end{deluxetable*}

\subsection{Comparison to Previous Orbits}

\begin{figure*}
\centering
\includegraphics[scale=0.9, trim=10mm 0mm 0mm 0mm]{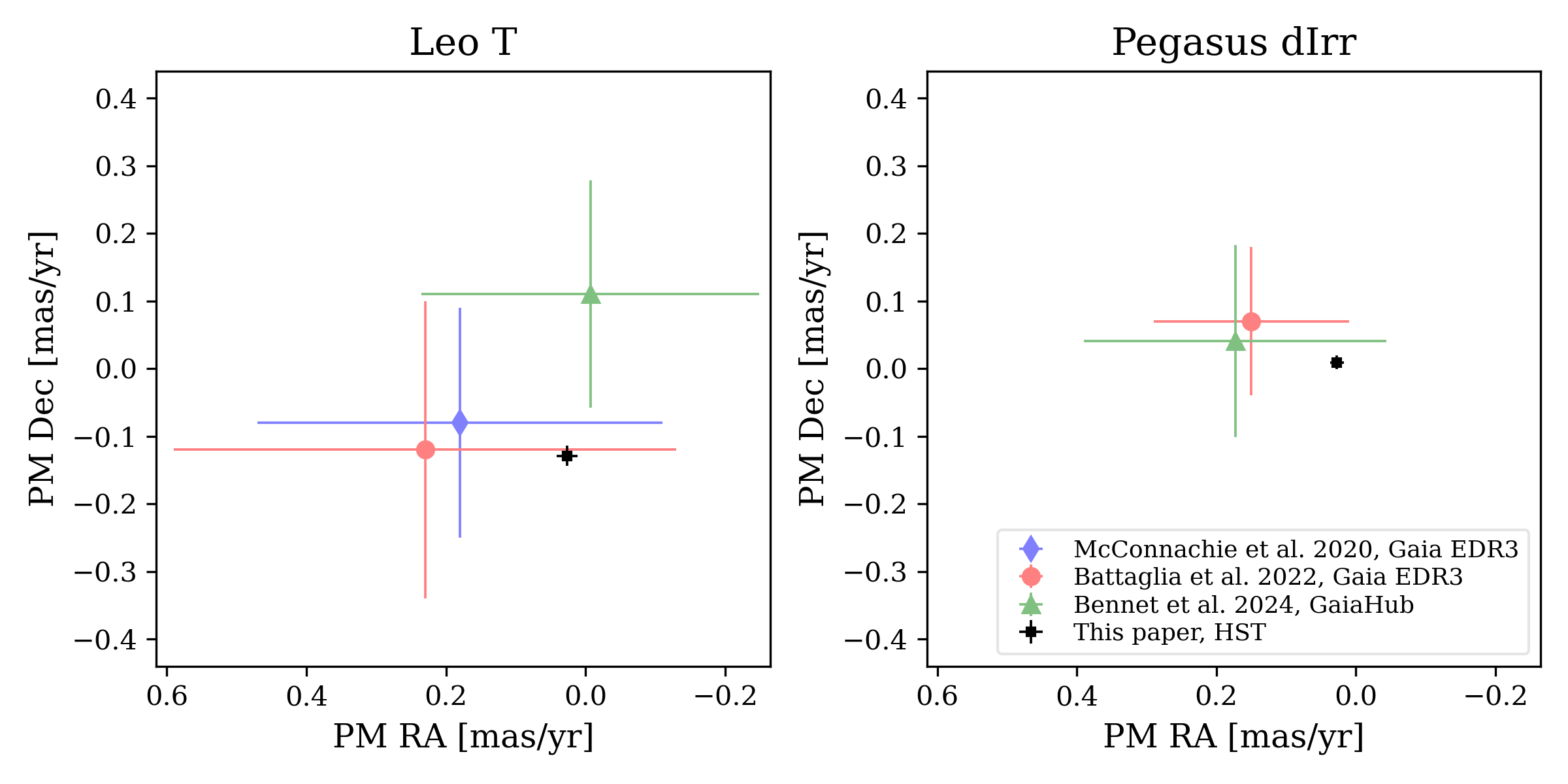}
\caption{Comparison of the absolute PM measurements for Leo~T and Pegasus, the only two dwarfs in our sample with existing literature PMs. We show \citet{McConnachie_2020b} in blue diamonds, \citet{Battaglia_2022} in red circles, \citetalias{Bennet_2024} in green triangles and our own results as black squares. 
As can be seen the PMs reported in this work are consistent with, though far more precise than, previous estimates.
The different values reported using \textit{Gaia} DR3 data are primarily the result of the different membership selection methods used by those works.}
\label{Fig:PM_comp}
\end{figure*}

Orbits for two galaxies in the present sample, Leo T and Pegasus, were previously published in \citetalias{Bennet_2024}. Here, we briefly compare the PMs and subsequent orbital histories between \citetalias{Bennet_2024} and this work. 

For both objects, we report substantial improvements in the uncertainty here. This is partly because the time baseline is substantially larger, increasing from 1.84~yrs to 8.01~yrs for Pegasus, and 4.37~yrs to 11.03~yrs for Leo~T. 
However, the greatest improvement is in the number of sources considered. In \citetalias{Bennet_2024}, the absolute reference frame is set by the stars in the \textit{Gaia} catalog; however, in this work, the reference frame is set by background galaxies. For these dwarfs, there are substantially more background galaxies in the \textit{HST} imaging than there are stars in the \textit{Gaia} catalog, yielding a more certain anchor to the absolute reference frame. 
Finally, a larger number of sources also allows a more robust selection of member stars compared to background objects, lowering potential contamination and improving the measured uncertainties. This improvement also applies to previous \textit{Gaia} measurements of Leo~T and Pegasus' PMs \citep[][]{McConnachie_2020b,Battaglia_2022} for similar reasons. Figure \ref{Fig:PM_comp} clearly illustrates the past PM measurements compared to our much improved PM measurements. 

\subsubsection{Leo T}
The Leo T velocity vector in \citetalias{Bennet_2024} derived from PMs was 

\[ v_{B24}= (-90\pm369, 427\pm305, 83\pm334) \rm \, km \,s^{-1}, \]
and here, we find 
\[ v = (34\pm23, 0\pm31, -66\pm22) \rm \, km \, s^{-1}\]

Not only is there a significant decrease in total velocity from 443$\pm$306 km s$^{-1}$ to only 86$\pm$16 km s$^{-1}$, but the component-by-component uncertainties are reduced by an order of magnitude. Despite these vastly different 3D velocity vectors, this still amounts to a first infall orbit around both the MW and M31. However, the estimated distance between Leo T and either the MW or M31 at 6 Gyr ago is much lower than previously found. 

From Tables \ref{tab:orbit_params_mw} and \ref{tab:orbit_params_m31} and the corresponding tables in \citetalias{Bennet_2024}, zero Leo T orbits relative to the MW contained either a pericenter or apocenter. In this work, we find that only one orbit (0.1\%) has an apocenter about the MW at 6 Gyr ago at very large distance ($\sim$930 kpc). 

On the other hand, there is a more substantial difference in the orbits for Leo T relative to M31. In \citetalias{Bennet_2024}, about 7\% and 1\% of Leo T orbits contained a pericenter and apocenter, respectively. Statistically, if a pericenter existed, it was most often in the range of $\sim$900-1000 kpc, and similarly between $\sim$1000-1200 kpc for apocenter. Pericenters were usually within the last 1 Gyr, and apocenters could occur in a wide range between $\sim$1-5 Gyr ago. By contrast, in this work, only two Leo T orbits (0.2\%) have a pericenter, but 16.1\% have apocenters. The two recovered pericenters occur at $\sim$6 Gyr ago and at distances of $\sim$1390-1530 kpc. Apocenter is constrained to distances of $\sim$1140-1260 kpc relative to M31 between 4.3-5.6 Gyr ago. 
We conclude Leo T is neither a satellite of the MW nor M31 nor a backsplash galaxy. The most significant differences compared to our previous work are that with the improved 3D velocity presented here, we can constrain orbital parameters, specifically apocenter, to a range of $\Delta(D_{M31})=$ 120 kpc.

\subsubsection{Pegasus}
In \citetalias{Bennet_2024}, the Pegasus velocity vector was 
\[v_{B24}= (-721\pm880, -64\pm506, 15\pm503)\, \rm km \, s^{-1}\]

and here, we find 
\[v= (-91\pm41, 93\pm29, 122\pm30) \, \rm km \,s^{-1}.\]

Similar to Leo T, the total 3D velocity of Pegasus decreases significantly from 724$\pm$693 km s$^{-1}$ to 179$\pm$41 km s$^{-1}$. Additionally, the uncertainties of the velocity vector are reduced by an order of magnitude. 

In \citetalias{Bennet_2024}, Pegasus was concluded to be a first infall galaxy relative to the MW. While about 16\% and 6\% of Pegasus orbits had a pericenter or apocenter, respectively, the distances for both orbital parameters were typically $>$ 850 kpc from the MW. In this work, these percentages increase to $\sim$78\% (pericenter) and $\sim$96\% (apocenter), yet the distances still typically remain significantly outside of the MW's virial radius ($\gtrsim$ 600 kpc). The backsplash probability is still non-zero at 0.8\% (see Table \ref{tab:splash}), meaning that a small fraction of orbits do reach within M31's virial radius.

On the other hand, Pegasus showed a 19.5\% chance of a pericenter within the virial radius of M31 at distances of $\sim$270-470 kpc in \citetalias{Bennet_2024}. Thus, a recent interaction with M31 was not ruled out, finding a 5.2\% probability of being an M31 backsplash galaxy. Here, zero orbits have a pericenter about M31 and only 0.4\% contain apocenters. However, these, too, are much larger than M31's virial radius, and therefore, Pegasus is not an M31 backsplash galaxy.

\section{Discussion} \label{sec:discussion}

\begin{figure*}
\centering
    \includegraphics[width=\textwidth]{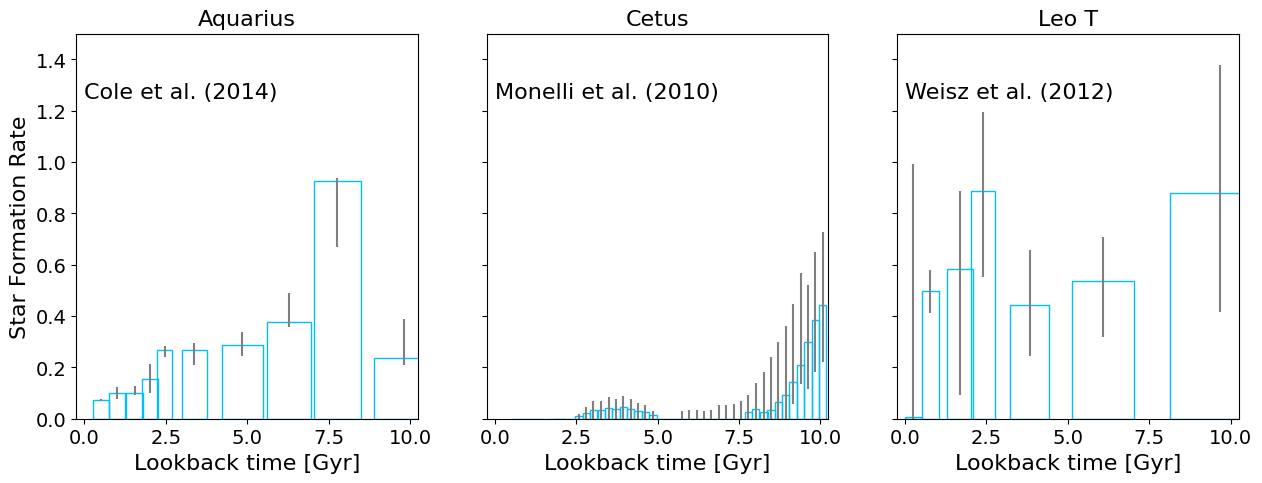}
    \caption{Star formation histories for the past 10 Gyr for an illustrative subsample of our dwarfs. The y-axis shows the star formation rate in arbitrary units. Uncertainties are shown in the gray lines. 
    Left panel: Aquarius from \citet{Cole2014}, Aquarius shows relatively constant star formation between 2.5 and 7 Gyr ago, with a major burst at just beyond $\sim$7~Gyr ago.
    Center panel: Cetus from \citet{Monelli_2010}, Cetus quenches at $\sim$8 Gyrs, the slight star formation around $\sim$4 Gyrs is likely caused by blue straggler stars. 
    Right panel: Leo~T from \citet{Weisz_2012}, Leo~T shows relative constant star formation similar to Aquarius, but significantly more stochastic. }
    \label{fig:SFH}
\end{figure*}

In the following section, we examine our dwarf sample galaxy by galaxy, making specific comparisons to previous literature regarding their natures and evolutionary histories. For illustrative purposes, Figure \ref{fig:SFH} shows the literature SFHs for Aquarius \citep{Cole2014}, Cetus \citep{Monelli_2010}, and Leo~T \citep{Weisz2014}.

\subsection{Aquarius}
\label{subsec:aquarius}

Aquarius is a very isolated dwarf at the edge of the LG, more than 1~Mpc from the MW or M31. It shows an SFH with a relatively constant star formation (see Figure \ref{fig:SFH}) between 2.5 and 7 Gyr ago, and a slight decline in the star formation rate in recent times (this may be a result of increased time resolution in the SFH toward the present). Its most notable SFH feature is a large star formation burst at $\sim$7~Gyrs ago \citep[][]{Cole2014}. Aside from this burst, which we will discuss below, this SFH is consistent with those of isolated dwarf galaxies in Aquarius' mass range, as found in simulations \citep[][]{Garrison-Kimmel2019}.

Aquarius also has a large neutral hydrogen reservoir with a gas mass of $M_{HI}=4.1\pm0.3\times10^{6} \, \rm M_{\odot}$, \citep[][]{Putman_2021} giving it a gas/stellar mass ratio of $M_{HI}/M_* \sim 1.5$, typical for an isolated dwarf galaxy \citep{huang2012}. 

Despite its isolation at the edge of the LG, the burst of star formation $\sim$7~Gyrs ago may be due to an interaction with a more massive host, but our orbits exclude the possibility of any close interactions between Aquarius and the MW or M31. This starburst also coincides with a slight dip in the average metallicity of stars produced in this epoch compared to earlier stars, as well as a larger metallicity spread for stars older than 8~Gyrs \citep{Cole2014}. One possibility inferred from the combination of observed phenomena around $\sim$7~Gyrs ago is a dwarf-dwarf merger, such that Aquarius merged with another, smaller dwarf galaxy around this time.  

In \citetalias{Bennet_2024}, we discussed evidence of dwarf-dwarf mergers in Leo~A ($\sim$6 Gyrs ago) and Leo~I ($\sim$5 Gyrs ago) due to similar peaks in their SFHs that cannot be explained by interactions with the MW or M31. Similar star formation bursts pointing to possible merger events are also seen in Fornax dSph \citep[][]{del_Pino_2015,Rusakov_2021} and Andromeda~II \citep[][]{del_Pino_2017}. However, in these cases, it is harder to separate the influence of the massive host (the MW or M31) from possible merger signals. Fornax dSph has been a long-term member of the MW satellite system \citep[][]{Fritz_2018,Rusakov_2021,Battaglia2022}, and therefore has a more complicated interaction history with the MW that cannot be ignored. There are currently no published PMs for Andromeda~II that would allow the kind of kinematic modeling needed to eliminate the possibility of the effects being caused by M31 or M33. 
If true, our results increase the number of possible dwarf-dwarf mergers in the LG at intermediate times to three instances. 

Other explanations for the star formation peaks include the inflow of pristine gas from the cosmic web or interactions with large-scale structure filaments, which could also trigger star formation bursts \citep[][]{Ledinauskas2018}. Purely internal dynamical processes could also explain the peaks \citep[][]{Cole2007}. While the metallicity features in these three dwarfs seem to favor a dwarf-dwarf merger hypothesis \citep[][]{Cole2014,Kirby2017,Ruiz-Lara_2021}, more data is needed to draw firmer conclusions.   

Simulations have shown that dwarf-dwarf mergers at intermediate times (4-8~Gyr)---such as our three possible mergers---occur from dwarf pairs that form and become bound together at early times (10-12 Gyrs) \citep{Chamberlain2024a,Chamberlain2024b}. Including our sample, there are 15 LG dwarfs in the literature with kinematics sufficiently precise to distinguish between a dwarf-dwarf merger and the influence of a massive host galaxy \citep[e.g.][]{Sohn_2013,Sohn_2017,Sohn_2020,McConnachie_2021,Battaglia2022,Bennet_2024}. This implies a potential merger fraction of $\sim$20\%, which matches that found in \citet{Chamberlain2024a,Chamberlain2024b} for low mass dwarfs. However, the mass range of the LG dwarfs examined in \citet{Chamberlain2024a,Chamberlain2024b} is lower, and the LG is a more crowded environment than those probed in those studies.

We also find that compared to the $\sim30$ classical dwarfs in the modern LG ($M_{V}\leq-8$ mag), we would expect to find $\sim3$ pairs at z=0 based on the above work and a $\sim$20\% merger fraction. If we compare this to the actual LG we find that there are fewer pairs than expected, with the LMC and SMC  being the only kinematically confirmed pair in this mass range\footnote{Other possible kinematic associations of dwarf galaxies in the MW halo such as the Leo-Crater group \citep[][]{Julio_2024} and the pair Pegasus~III and Pisces~II \citep[][]{Richstein_2022} have been reported using 3D kinematics. However, these galaxies are ultra-faint dwarfs rather than the classical mass dwarfs as those explored in this work.}. More work is needed to understand the frequency of dwarf-dwarf mergers across a wider range of dwarf masses in the LG.

\subsection{Cetus} 
\label{subsec:cetus}
Cetus and Tucana have been used extensively as examples of so-called backsplash galaxies, i.e., galaxies that are outside the virial radius of a more massive galaxy or cluster at present, but retain evidence of a previous interaction \citep[e.g.][]{Teyssier_2012,Taibi_2018,Blana2020,Santos_santos_2023}. 

More generally, this definition, in part, is also motivated by a lack of significant amounts of HI gas or recent star formation (see Figure \ref{fig:SFH}). Cetus' SFH shows substantial star formation after reionization, however, this stops around $\sim8$~Gyrs ago as Cetus undergoes rapid quenching \citep{Monelli_2010}. 
Cetus and Tucana are the only isolated LG dwarfs that exhibit these properties, separating them from the rest of the isolated LG dwarfs, which are either transitional or irregular dwarfs, with evidence of recent star formation and HI reservoirs \citep{Grcevich_2009,Spekkens_2014,Putman_2021}.

Furthermore, our GBT observations of Cetus (see Section \ref{subsec:gbt}) were a non-detection pushing the HI upper limit for this galaxy down to $8\times10^{4}\, \rm M{_\odot}$ from the previous upper limit reported in \cite{Putman_2021} ($9.6\times10^{4} \, \rm M{_\odot}$). Therefore, we conclude that Cetus is a gas-poor dwarf with an HI upper limit of $\sim$3\% of the reported stellar mass \citep{McConnachie_2012}. This is the most robust HI gas limit in the LG outside the MW halo. 

Though many studies have labeled Cetus as a backsplash galaxy based on these properties, some previous kinematic studies have indicated that these galaxies are more likely to be truly isolated \citep{Fraternali_2009,Buck_2019}. These works examine the radial velocity and distance of Cetus and Tucana from the M31 and the MW, and compare this to simulated dwarfs in LG analogs to explore the probability that a dwarf with the observed properties is either backsplash or on first infall. These show that dwarfs with Tucana and Cetus' properties are likely (but not certain) to be on first infall.

In Section \ref{sec:Orbits}, we showed that Cetus only has a low probability of being a backsplash galaxy of the MW or M31 within the last 6 Gyrs (see also Table \ref{tab:splash}). While our orbits cannot accurately account for early universe interactions ($>$6 Gyr ago), the possibility that Cetus interacted with either massive galaxy beyond 6 Gyr ago may be higher than discussed thus far.

\begin{figure*}
    \centering
    \includegraphics[width=1\linewidth, trim=5mm 5mm 0mm 0mm]{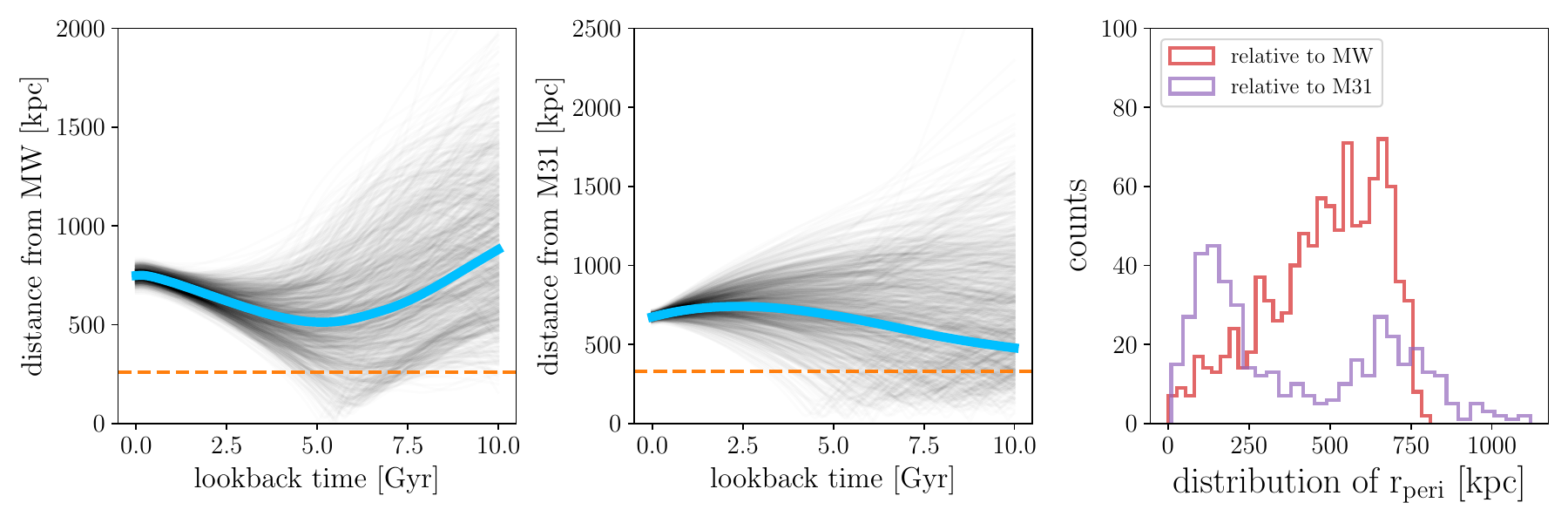}
    \caption{The left and middle panels are similar to the column corresponding to Cetus in Figure \ref{fig:errors}, except the integration period is extended back to 10 Gyr ago. The right panel shows the distribution of distances at pericenter ($\rm r_{peri}$) relative to the MW and M31 in the orbits where one exists. Close passages (i.e., within the virial radius of the MW or M31) are more common if Cetus is a backsplash galaxy of M31.}
    \label{fig:cetus_10Gyr}
\end{figure*}

For illustrative purposes, we integrated both the direct orbit and orbital uncertainties for Cetus back to 10 Gyr ago, extending the previous integration period. Figure \ref{fig:cetus_10Gyr} shows the resulting orbital histories with respect to the MW in the left panel and M31 in the middle panel. In the last 10 Gyr, 100\% of Cetus orbits have a pericenter around the MW at about 6 Gyr ago. Furthermore, Cetus could have interacted with M31 between 5-10 Gyr. When we recompute the backsplash probability for Cetus, the likelihood increases to 13.2\% relative to the MW and 22.9\% relative to M31. 

The right most panel of Figure \ref{fig:cetus_10Gyr} shows the distribution of $\rm r_{peri}$, the distance at pericenter, relative to the MW and M31, corresponding to the orbits in the left and middle panels. These distributions indicate that passages around the MW were more likely to occur at large distances (i.e., two times the MW's virial radius, or $\approx$500 kpc). In contrast, there is a bimodal distribution in the $\rm r_{peri}$ distribution relative to M31 such that the first distribution peaks within M31's virial radius of 329 kpc. Therefore, while the likelihood of a passage around M31 is less statistically common (44\% of orbits), a close passage around M31 is more likely. Additionally, the burst of star formation around 8 Gyr ago is more consistent with the timing of a passage around M31 ($1\sigma$ for $\rm t_{peri}=5.32-8.94$ Gyr ago. Together, our orbital results combined with the Cetus SFH paint a consistent picture that it could be a backsplash galaxy of M31. 
Our conclusions are consistent with those of simulations \citep{Garrison-Kimmel2019,Joshi_2021,Samuel_2022} showing that a dwarf with the mass of Cetus should not self-quench without the influence of a massive host \citep{Garrison-Kimmel2019}; though it readily quenches if such a host is present. 

We emphasize that this exercise is meant to demonstrate possible orbits for Cetus in the early Universe; however, due to the assumptions included in our rigid potential models, we do not suggest interpreting definitive conclusions about Cetus' past orbital history from this analysis. More accurate PMs and improved orbital models will be required to limit the possible parameter space for Cetus, but we conclude that the quenched nature of Cetus could be explained through an earlier interaction with one of the massive LG galaxies. Fortunately, HST Cycle 33 18032 will obtain fourth epoch data for Cetus, increasing the time baseline between the first and final epochs from 14 years to 20 years. With this forthcoming data, we may be able to robustly constrain Cetus' interaction history at earlier times (i.e., $>$ 6 Gyr ago)\footnote{As no additional epochs of data are expected for the remaining dwarfs in our sample, we refrain from illustrating orbits to 10 Gyr ago for the aforementioned reasons.}.

A final possibility for Cetus if its quenching is not a result of an interaction with the MW or M31 is that it was once part of a group infall into the LG with the LMC, M33, or another LG galaxy that came in earlier than 6 Gyr ago, and was then tidally removed from its companion(s), leaving Cetus behind. Interactions during this scenario may plausibly explain Cetus' SFH. Another consideration is that halo deformations due to the passage of massive satellites like the LMC and subsequent center of mass responses from massive host galaxies are not accounted for with the rigid potentials assumed in this work \citep[see][]{GaravitoCamargo_2021,Patel_2025}. This phenomenon is beyond the scope of this analysis, but it is important to acknowledge that changing halo shape can alter orbital histories.

\subsection{Leo T}
\label{subsec:leot}
Leo~T is the lowest stellar mass galaxy with a confirmed HI reservoir. This HI reservoir is symmetric, with a low velocity dispersion disk \citep[][]{Adams_2018}, supporting a history of non-interaction. Its SFH shows relatively constant star formation over the past $\sim$8~Gyrs \citep[][]{Weisz_2012,Clementini_2012}. These studies also show possible cessation of star formation in Leo~T in the past $\sim$25~Myrs, though this may be caused by the stochastic sampling of the initial mass function in such a low-mass galaxy (see Figure \ref{fig:SFH}). 

Leo~T has a notably low 3D velocity (see Section \ref{sec:Orbits} \& Figure \ref{fig:escape}), which is intriguing as it does not appear to be at apocenter, where a low velocity would typically be expected. Instead, it is currently accelerating as it moves toward the MW (see top panels of Figure \ref{fig:orbits}). This motivates acquiring PMs for other dwarf galaxies at similar radii relative to the MW and M31, e.g., Phoenix or Eridanus II, to understand whether the kinematics of Leo T is an anomaly or common amongst dwarfs in similar environments.

\subsection{Pegasus}
\label{subsec:pegasus}
Pegasus' recent ($\leq$6~Gyr) SFH shows an approximately constant star formation rate \citep[][]{Savino_2025}. 
This constant star formation rate also matches those of isolated dwarfs of Pegasus' mass in simulations \citep[][]{Garrison-Kimmel2019}. Our orbits substantiate these ideas, showing a very slim (0.8\%) chance of Pegasus being a backsplash galaxy despite previous kinematic studies suggesting this was possible \citep[][]{Buck_2019}.

HI imaging of Pegasus shows an undisturbed solid-body-like HI disk \citep{Kniazev2009} whose rotation is consistent with its RGB stars \citep{Higgs2021}, as expected for a classical dwarf that has had no interaction with a massive host.

\subsection{Pisces} \label{subsec:pisces}

Pisces is the only dwarf in the present sample that resides within the virial radius of the MW or M31. At present, it is just inside of M31's virial radius, having arrived there in just the last $\sim$100 Myr. Pisces has a small, disordered HI reservoir \citep[][]{Hunter_2012} and an SFH that shows significant mass assembly in the early Universe, followed by a slow decline for most of the past 10~Gyrs \citep[][]{Hidalgo_2011}. The exception to this is a large burst of star formation about 1.5~Gyrs ago; this burst does not correlate with any encounter with M31 based on our orbital histories.

We concluded Pisces (LGS~3) is likely to be on first infall into M31's halo, however, there has been much discussion historically about whether Pisces is a member of the M31 or M33 systems \citep[][]{Lee_1995,Aparicio_1997,Cook_1999,McConnachie_2005,McConnachie_2012}.

Recent RR Lyrae distance estimates \citep[][]{Savino_2022} move Pisces significantly closer ($\sim$150~kpc) to the MW compared to previous TRGB estimates \citep[][]{McConnachie_2012}. The new 3D separation between Pisces and M31 (260~kpc) is similar to that between Pisces and M33 (235~kpc). In a companion paper, we will explore the recent interaction history between Pisces and M33 in the last $\sim$2 Gyr and how the categorization of Pisces as an associated dwarf of M31 versus M33 impacts interpretations of recent observational surveys.

\subsection{Tucana}
\label{subsec:tucana}

Tucana is commonly assumed to be a backsplash galaxy based on its SFH and HI gas properties. It has no detectable HI reservoir \citep[][]{Spekkens_2014, Putman_2021}, and has no star formation in the past 8~Gyrs after it quenched around this time \citep[][]{Monelli_2010b}. \cite{Fu_2024} also examined the metallicity distribution function of Tucana and compared it to other dwarfs in the LG. This showed that the metallicity gradient in Tucana is far more similar to those of the isolated galaxies of the LG (e.g., Leo~A) rather than those that have interacted with the MW or M31 (e.g., Draco dSph).

Unlike with Cetus, we do not find any evidence that Tucana has interacted with the MW or M31 in the last 6 Gyr (e.g., 0\% backsplash probability). When we integrate back further to 10 Gyr ago, the distance between Tucana and the massive galaxies continues to increase moving backward with time. Thus, our conclusion that Tucana is on its first passage around the MW and first infall around M31 remains consistent. 

Again, it is possible that Tucana was accreted into the LG $>$ 6 Gyr ago along with the LMC, M33, or another galaxy, became tidally stripped from its companion, which may have led to the cessation of SF, and has since evolved in relative isolation.

Simulated dwarfs with masses similar to that of Tucana ($1.1\times10^{6}\, \rm M_{\odot}$) suggest these galaxies should be quenched even if isolated from the influence of a massive host \citep{Digby_2019,Garrison-Kimmel2019,Joshi_2021}. However, Tucana's relatively low mass, compared to other dwarfs in the LG, lies in the regime where many zoom-in simulations suffer from resolution effects. The consequence of this is typically early quenching in almost all small dwarfs ($M_{*}\lesssim1\times10^{6}\, \rm M_{\odot}$; \citealt{Garrison-Kimmel2019}). This is in contrast to the LG, where dwarfs, such as Phoenix and Leo T, are still star-forming, despite having masses less than or equal to Tucana \citep[][]{Hidalgo_2009,Weisz_2012}. Thus, most simulated dwarfs and Tucana quench for very different reasons. 

Recently, the Lyra simulations \citep[][]{Gutcke_2022}, which simulated single isolated dwarfs to a mass resolution of individual stars, show that Tucana-mass dwarfs are at the boundary of being quenched by reionization. Some dwarfs in this mass range survive and continue to form stars until recent times, while others quench during reionization. However, these detailed simulations do not include any direct analogs for Tucana, where the dwarf survives reionization and then quenches later at intermediate times ($\sim$8~Gyrs). 

Therefore, the evolution of Tucana is hard to explain with our current models of dwarf galaxy evolution. Studies of other isolated, quenched dwarfs such as Hedgehog \citep{Li_2024} also speculate about whether Hedgehog is a backsplash galaxy of the Centaurus~A group. However, they also suggest other possible explanations for its quenching, including ram pressure stripping in the cosmic web \citep[see also][]{Benavides_2025}, reionization, or internal processes such as supernova and stellar feedback. The same mechanisms may play a major role in the quenching of Tucana. Observations of additional quenched field dwarfs are needed to disentangle these possible scenarios.

\section{Summary and Conclusion} 
\label{sec:conclusion}

In summary, we used \textit{HST} observations to measure the PMs of six isolated Local Group dwarf galaxies. Four of these are the first reported PMs, while the other two greatly improve the precision compared to earlier results. We use these results to report the current positions and velocities of the dwarfs in Galactocentric coordinates. 

We reconstruct orbital histories for the six dwarfs following the methods of \citetalias{Bennet_2024}, using a joint galaxy potential model that includes forces from the most massive halos in the LG, including the MW, M31, the LMC, and M33. All galaxies are modeled as extended but rigid mass distributions, and we track the simultaneous center of mass motion of all galaxies over time, as they experience the gravitational influence of each other. Below, we briefly summarize the conclusions of this analysis, which aims to provide constraints on first infall vs. backsplash (i.e., having a pericenter at a distance less than the virial radius of the MW or M31 in the last 6 Gyr) scenarios relative to the MW and M31.

\begin{enumerate}\setlength{\leftmargin}{0pt}
    \item Based on the Galactocentric coordinates derived in this world, all six dwarfs are bound to the Local Group (LG) as a whole but not necessarily to either the MW or M31 individually. This highlights the importance of modeling the LG potential rather than single halos in isolation.
    
    \item Cetus shows a small probability ($\sim$4–6\%) of being a backsplash galaxy relative to the MW or M31 in the last 6 Gyr. This probability increases when we integrate its orbit and uncertainties to 10 Gyr ago. While this is the regime where rigid orbital modeling becomes less reliable, a close ($\rm r_{peri} < R_{vir}$) passage relative to M31 may coincide with the burst of star formation seen in its SFH $\sim$8 Gyr ago. Improved PM measurements and more sophisticated orbital models are needed to constrain the history of Cetus further.
    
    \item Though commonly thought to be a backsplash galaxy due to its lack of HI reservoir and lack of recent star formation, our orbital results show Tucana has a 0\% probability of having passed within the virial radius of either the MW or M31 in the last 6 Gyr, ruling out a backsplash origin. Tucana very likely completed a pericenter around the MW at 2.3 Gyr ago, only reaching distances $\rm >2R_{vir,MW}$; therefore, we conclude Tucana is on its first passage. Other possible explanations for its quenching must be considered, including ram pressure stripping in the cosmic web, reionization, or internal processes such as supernova and stellar feedback.
    
    \item Pisces is currently a satellite of M31 (i.e., within its virial radius) and likely on its first infall; past pericenters are possible but rare ($\leq$1.3\%). A more detailed analysis linking Pisces to potential interactions with M33 and M31 will be discussed in a companion paper (Patel, Bennet, et al., in prep.).
    
    \item The remaining dwarfs (Aquarius, Leo T, Pegasus, Tucana) show no strong evidence for past interactions within the virial radii of MW or M31, and are therefore most consistent with first infall. Pegasus shows very low backsplash probabilities ($\leq$0.8\%), while Aquarius and Leo T have zero probability. Our results are consistent with ongoing or recent star formation expected for isolated, first-infall dwarfs.
\end{enumerate}

Star formation properties of the satellites in the SAGA survey, which took a census of satellite luminosity functions down to the magnitude of Leo~I around 101 MW analogs, also defined as those within the projected virial radii of their host galaxies \citep{Mao_2024, Geha_2024}.  \citep[][]{Mao_2024} shows that the majority are star-forming, particularly far from the host galaxies. However, we do see that $\sim$20\% are quenched even when at the virial radius of the host galaxy. This is also found in the ELVES sample \citep[][]{Carlsten_2022}. This implies that these dwarfs have quenched prior to encounters with the host or are backsplash galaxies. 

If we examine the LG as a whole, we find 10 dwarfs isolated from either the MW or M31 in this mass range. Among these, Cetus and Tucana are the only dwarfs that are quenched, with the other 8 being star-forming dwarf irregulars or transition dwarfs. This is consistent with the quenched dwarfs in SAGA and ELVES at the edge of the host's virial radius as being analogs to Cetus and Tucana, and therefore potentially quench prior to interaction with their hosts. Similarly, simulations \citep[e.g.,][]{Garrison-Kimmel2019,Joshi_2021,Samuel_2022} typically quench dwarfs like Cetus and Tucana via host interactions, but this study’s results suggest alternative or earlier quenching mechanisms must be considered (e.g., reionization, internal feedback, dwarf-dwarf mergers).

This work emphasizes the need to explore a range of quenching mechanisms for intermediate mass ($M_* \approx 10^5–10^7 \, \rm M_{\odot}$) dwarfs. While the PM measurements and orbital reconstructions reported in this work provide essential benchmarks for future studies of star formation, quenching, and environmental influence on dwarf galaxies in the LG, current orbital models are also limited in scope.

For example, the halos of all galaxies included in the orbital models are rigid, and therefore any deformation of the dark matter halos of the MW and M31, in particular, due to the passage of satellites is not included \citep[e.g.,][]{GaravitoCamargo_2021}. As shown in \citet{GaravitoCamargo_2021}, the outskirts of the MW's halo become deformed out to the virial radius. In forthcoming work (Patel, Garavito-Camargo et al., in prep.), we will explore the impact of this deformation on the orbits of LG dwarf galaxies.

\begin{acknowledgments}
Support for this work was provided by NASA through grants for program GO-15911 and GO-17174 from the Space Telescope Science Institute (STScI). STScI is operated by the Association of Universities for Research in Astronomy, Incorporated, under NASA contract NAS5-26555. EP is financially supported by NASA through the Hubble Fellowship grant \# HST-HF2-51540.001-A awarded by the  STScI. This work was based on observations taken with the NASA/ESA HST and obtained from the Data Archive at STScI. AdP acknowledges financial support from the \emph{Severo Ochoa} grant CEX2021-001131-S (MICIU/AEI/10.13039/501100011033), the Ram\'on y Cajal fellowship RYC2022-038448-I (MICIU/AEI/10.13039/501100011033, co-funded by the European Social Fund Plus), the Spanish Ministry of Science and Innovation project PID2021-124918NB-C41 (MICIU/AEI/10.13039/501100011033, FEDER, EU), and the RyC-MAX grant 20245MAX008 (CSIC).
\end{acknowledgments}

All the {\it HST} data used in this paper can be found in MAST: \dataset[10.17909/3r3f-9j87]{http://dx.doi.org/10.17909/3r3f-9j87}.

\software{Numpy \citep{numpy},
  SciPy \citep{SciPy-NMeth},
  Matplotlib \citep{matplotlib},
  IPython \citep{ipython},
  Jupyter \citep{jupyter}, 
  Astropy \citep{astropy:2013, astropy:2018, astropy:2022}, 
  Gala \citep{gala}}

\bibliography{refs}{}
\bibliographystyle{aasjournalv7}

\end{document}